\documentclass[aps,twocolumn,pra,groupedaddress,showpacs,floatfix,superscriptaddress]{revtex4}

\usepackage{graphicx}
\usepackage{dcolumn}
\usepackage{bm}
\usepackage{amssymb}
\usepackage{amsmath}
\usepackage{color}
\usepackage[normalem]{ulem}
\usepackage{subfigure}

\usepackage{ae,aecompl}
\usepackage{verbatim} 
\usepackage{cancel}
\usepackage[T1]{fontenc}
\usepackage[ansinew]{inputenc}
\usepackage{cancel}
\usepackage{tabularx}                   
\usepackage{sidecap}                    

\begin{document}

\preprint{APS/123-QED}

\title{Sub-milliKelvin spatial thermometry of a single Doppler cooled ion in a Paul trap}

\author{S.~Kn\"{u}nz}
\affiliation{Max--Planck--Institut f\"ur Quantenoptik, 85748
Garching, Germany}
\author{M.~Herrmann}
\affiliation{Max--Planck--Institut f\"ur Quantenoptik, 85748
Garching, Germany}
\author{V.~Batteiger}
\affiliation{Max--Planck--Institut f\"ur Quantenoptik, 85748
Garching, Germany}
\author{G.~Saathoff}
\affiliation{Max--Planck--Institut f\"ur Quantenoptik, 85748
Garching, Germany}
\author{T.~W.~H\"ansch}
\affiliation{Max--Planck--Institut f\"ur
Quantenoptik, 85748 Garching, Germany}
\author{Th.~Udem}
\affiliation{Max--Planck--Institut f\"ur Quantenoptik, 85748
Garching, Germany}

\date{\today}

\begin{abstract}
We report on observations of  thermal motion of a single,
Doppler-cooled ion along the axis of a linear radio-frequency
quadrupole trap. We show that for a harmonic potential the thermal
occupation of energy levels leads to Gaussian distribution of the
ion's axial position. The dependence of the spatial thermal spread
on the trap potential is used for precise calibration of our imaging
system's point spread function and sub-milliKelvin thermometry. We
employ this technique to investigate the laser detuning dependence
of the Doppler temperature.
\end{abstract}


\maketitle

\section{Introduction}

In the final stages of laser cooling the motion of an atom is
dominated by the random recoils of photon absorption and emission
events~\cite{wineland,stenholm}. If the atom is harmonically
confined, this Brownian motion~\cite{brown,einstein} is expected
to result in a Gaussian distribution of its position and
velocity~\cite{uhlenbeck,blatt,meekhof}. The width of this
distribution can be intuitively interpreted as temperature, which we
define as the time averaged energy divided by Boltzmann's
constant for a single particle. This is a powerful notion, since
many experiments require low residual kinetic energy, e.g., for
precision
metrology~\cite{alclock,maxmg,forcedetection,injkn,ionlockin} or
quantum computation and simulation~\cite{qsim,qcomputer}. In this
article we study the spatial probability density of a Doppler cooled
Mg$^+$ ion trapped in a linear radio frequency (rf) trap, confirm
the expected Gaussian distribution, and demonstrate that our
straightforward imaging approach enables precise thermometry, as
required for a wide range of experiments.

While in the strong-binding limit the comparison of the strengths of
motional sidebands allows precise temperature
measurements~\cite{bergquist,slodicka}, in the weak-binding limit
the sidebands are not resolved. In this regime, ion temperatures are
usually derived from the fluorescence line shapes which are
decomposed into their Lorentzian lifetime contribution and the
thermal distribution~\cite{liftemperature} by fitting a Voigt
function. However, this method relies on the assumption of a
Gaussian thermal distribution and the separation of the lifetime and
thermal widths can be accompanied by rather large
uncertainties~\cite{hasegawa,maxmg}. Upper limits for the
temperatures of cooled ions have also been obtained by measuring the
(thermal) spatial distribution in early laser cooling
experiments~\cite{neuhauser}. Uncertainties down to 5~mK have
recently been reported~\cite{kielpinski} by means of a specifically
designed Fresnel lens with high spatial resolution. A similar
technique has also been applied to atoms~\cite{dotsenko}.
Thermometry on large ion crystals has been performed by comparing
crystal images to the results of molecular dynamics
simulations~\cite{drewsen,roth}.

\begin{figure}[tb]
        \centering
        \includegraphics[width=0.85\columnwidth]{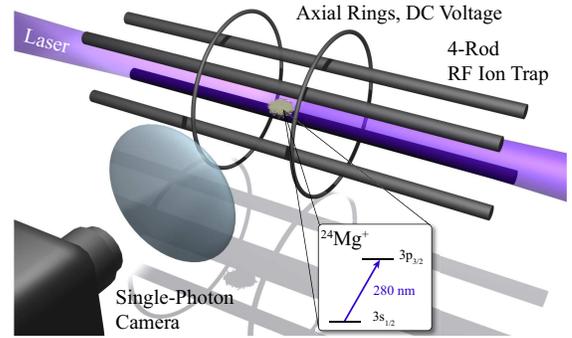}
    \caption{(Color online) Experimental setup: A $^{24}$Mg$^+$ ion is trapped in a linear
    Paul trap. Radial confinement is accomplished by an rf voltage applied
    to two of the four rods. A dc voltage applied to the rings provides tunable
    axial confinement. The ion is Doppler cooled by red-detuned laser
    light addressing the cycling  D$_2$ transition. A single-photon
    camera observes the axial spatial fluorescence distribution via an imaging system.
        }
        \label{fig:setup}
\end{figure}


In this work (see also~\cite{sebphd}), we investigate the
time-averaged spatial distribution of a single Mg$^+$ ion confined
in a linear quadrupole trap and laser-cooled {close} to the Doppler
limit. The trap is operated with weak axial dc confinement
which results in an axial spatial spread considerably larger than
the resolution of our imaging optics. As expected for Doppler
cooling, we observe a Gaussian fluorescence distribution. By
accurately calibrating both the magnification and the resolution of
our imaging optics, we are able to measure accurate values of the
thermal spread which allow to extract ion temperatures with sub-mK
uncertainties. We employ this precise thermometry to investigate the
laser detuning dependence of the Doppler temperature.


\section{theoretical background}

We consider an ion of mass $m$, trapped along the $z$-axis in a
harmonic potential \mbox{$V=m\omega^2 z^2/2$} with secular frequency
$\omega$, leading to motional quantum mechanical oscillator
states $\psi_n(z)$ of energy \mbox{$E_n=(n+1/2)\hbar\omega$}. Under
the assumption that the ion's random walk caused by the stochastic
photon absorption and emission is ergodic, the single ion can be
assigned a temperature $T$ which quantifies the time-averaged
occupation number of the states \mbox{$\overline
n=\left(\exp[\hbar\omega/k_bT]-1\right)^{-1}$}~\cite{reif} with the
Boltzmann constant $k_b$. The population probabilities $P_n$ of the
states $\psi_n(z)$ follow the distribution \mbox{$P_n=\overline
n^n\left(\overline n +1\right)^{-\left(n+1\right)}$}~\cite{reif}
which translates into a time-averaged spatial distribution of the
ion around its mean position $z_0$
\begin{align}
        f(z)=\sum_{n=0}^{\infty} P_n |\psi_n(z)|^2=\frac{1}{\sqrt{2\pi}\Delta z_{th}} e^{-\frac{(z-z_0)^2}{2\Delta z_{th}^2}}.
        \label{eq:pz}
\end{align}

In the evaluation of the sum we used Mehler's Hermite polynomial
formula~\cite{Watson}.
The variance of this Gaussian is \mbox{$\Delta
z_{th}^2=(\overline{n}+1/2)\hbar/(m\omega)$}. For $k_bT\gg
\hbar\omega$ as appropriate for the weak binding regime we have
$\overline{n}\approx k_bT/(\hbar\omega)$ and the root mean square
(RMS) width
\begin{align}
         \Delta z_{th}\approx\sqrt{\frac{k_bT}{m\omega^2}}.
        \label{eq:zrms}
\end{align}

\begin{figure}[t]
        \centering
        \includegraphics[width=0.85\columnwidth]{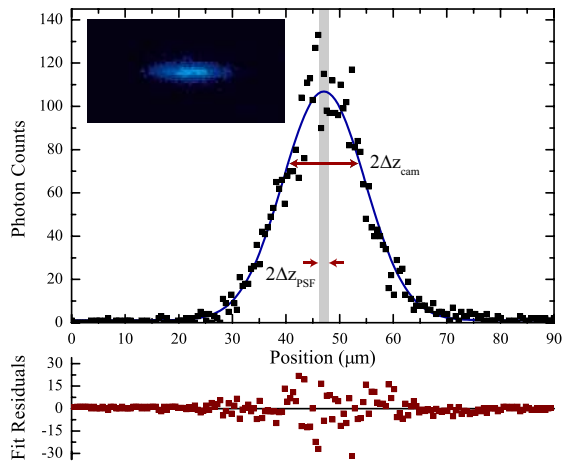}
    \caption{(Color online) Time-averaged spatial distribution of a
    single laser-cooled ion trapped with secular frequency
    $\omega=2\pi\times 15$~kHz ($T\approx1$~mK).  The inset shows the ion image while the plot
    is a histogram in axial direction. The residuals (lower graph) of a Gaussian
    fit (solid line) confirm a normal distribution as expected from Brownian motion
    caused by the recoils of stochastic photon scattering events.
    The RMS width $\Delta z_{cam}$ of the Gaussian and the RMS resolution \mbox{$\Delta z_{PSF}\approx1$~$\mu$m} of our
    imaging system are indicated.}
        \label{fig:spreadw}
\end{figure}

The temperature limit of a Doppler-cooled ion due to
secular motion in a harmonic potential results from an equilibrium
of laser-induced cooling and heating rates and is given
by~\cite{leibfried}

\begin{align}
        T= \frac{\hbar \Gamma}{8k_b}(1+\xi) \left((1+s)\frac{\Gamma}{2|\Delta|} + \frac{2|\Delta|}{\Gamma}\right).
        \label{eq:temperature}
\end{align}

It depends on the laser detuning $\Delta <0$ and the laser intensity
\mbox{$I=sI_{sat}$}, where $s$ and  $I_{sat}$ are the saturation
parameter and intensity, respectively. $\Gamma$ is the natural
linewidth of the optical dipole transition and $\xi=2/5$ takes the
dipole emission pattern into account. The temperature diverges for
\mbox{$|\Delta|\rightarrow0$}, \mbox{$|\Delta|\rightarrow\infty$},
and \mbox{$s\rightarrow\infty$}. The minimum of
\mbox{$T_{min}=\sqrt{1+s}(1+\xi)\hbar\Gamma/4k_b$} is obtained at a
detuning of \mbox{$\Delta_{min}=-\Gamma\sqrt{1+s}/2$} which, for
small $s$, reduces to $\Delta_{min}\approx -\Gamma/2$.

\begin{figure*}[t]
        \centering
        \subfigure[]{\includegraphics[width=0.9\columnwidth]{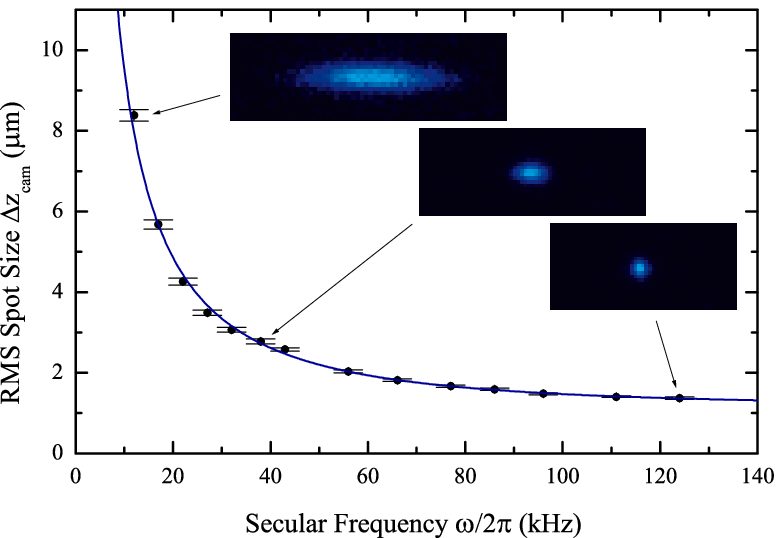}}
        \quad
        \quad
        \quad
        \subfigure[]{\includegraphics[width=0.89\columnwidth]{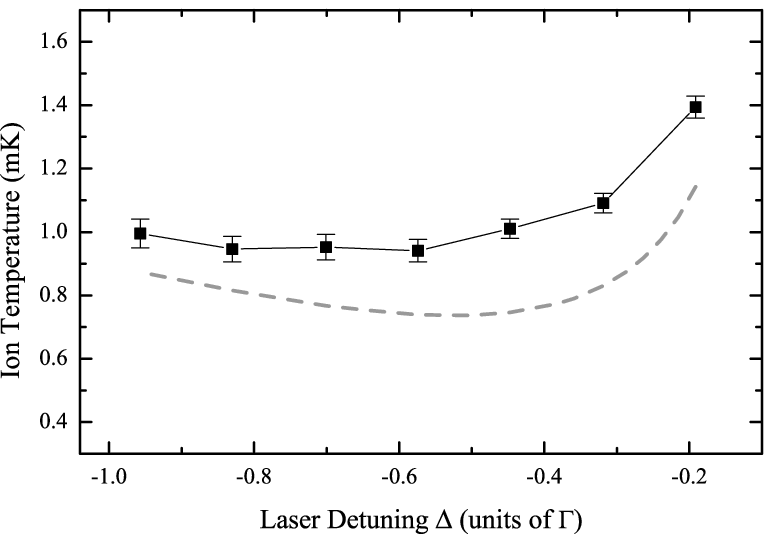}}
        \caption{(Color online) (a)~Measurement of the spot size $\Delta
        z_{cam}$ of the ion images with \mbox{$\Delta=-2\pi\times18.7$~MHz} and
        $s\lessapprox0.1$ for various trapping potentials $\omega/2\pi$. The
        insets show exemplary ion images. From a fit {(solid line)} of
        Eqs.~\ref{eq:zrms} and \ref{eq:width}, we obtain an ion temperature of
        $T=1.02(3)$~mK and a PSF of $\Delta z_{PSF}=1.13(3)$~$\mu$m RMS with
        statistical uncertainties. (b)~Same analysis for various laser detunings
        $\Delta$. Results are shown as squares together with temperatures expected
        from Eq.~\ref{eq:temperature} (dashed gray line). We attribute the observed
        temperature excess of up to $\approx0.2$~mK to systematic micromotion, see
        discussion in the text. The values of $\Delta
        z_{PSF}$ obtained for all data points agree within their statistical
        uncertainties. This measurement demonstrates the precision of the thermal
        spread thermometry providing a total accuracy of $<0.3$~mK unchallenged by
        other methods in the unresolved sideband regime.}
        \label{fig:iontemperature}
\end{figure*}

\section{Experimental Setup}

Our experimental setup is shown in Fig.~\ref{fig:setup} (see
also~\cite{injkn}). A single $^{24}$Mg$^+$ ion is trapped in a
linear rf quadrupole trap which operates at a trap frequency
$\Omega=2\pi\times 22.6$~MHz and generates radial rf confinement
with a secular frequency $\omega_r\approx2\pi\times 1$~MHz. The trap
electrodes are surrounded by two rings with a dc voltage applied to
generate axial confinement tunable from $\omega\approx2\pi\times 10$
to $150$~kHz. These frequencies are measured by secular excitation
with a weak external signal and can thus be controlled with an
accuracy of $\Delta\omega<2\pi\times 1$~kHz. Since
$\omega_r\gg\omega$, axial and radial motion are decoupled so that
we can neglect radial movement in the following. Due to its zero
nuclear spin, the alkali-like spectrum of $^{24}$Mg$^+$ shows no
hyperfine structure and the D$_1$ and D$_2$ lines constitute clean
cycling transitions well suited for Doppler cooling without the need
for repumper lasers. Two laser beams each stabilized in frequency
and intensity address the \mbox{$3^2S_{1/2}$-$3^2P_{3/2}$} D$_2$
transition near 280~nm (natural linewidth $\Gamma=2\pi\times
41.8(4)$~MHz~\cite{ansbacher}, $I_{sat}=2.50$~kW/m$^2$). {One beam
is aligned along the axial trap direction. The second laser beam is
slightly tilted by 14$^{\circ}$ against the first beam to provide
radial cooling.} The beams are detuned with respect to each other by
$\approx500$~kHz to avoid a stable interference pattern. The ion is
imaged with a $f/2$ condenser lens and a microscope objective onto a
single-photon camera (SPC, Quantar Mepsicron II). Because of the
SPC's limited spatial resolution of 56~$\mu$m, a magnification of
$M\approx100$ is chosen for the imaging system. The detector plane
is digitized in pixels of 49~$\mu$m size. The conversion factor
between the real space object size in $\mu$m and image size in
pixels is calibrated accurately by measuring the distance between
two simultaneously trapped ions in the imaging plane of the camera
for several trapping potentials $\omega$~\cite{james}. The RMS
resolution of the imaging system's point spread function (PSF) at
280~nm is \mbox{$\Delta z_{PSF}\approx 1$~$\mu$m} for optimal
alignment. It derives from the resolution of the lens system, the
resolution of the camera, and the discretization of the camera data.
However it is diminished when the ion is shifted out of focus,
particularly by varying laser forces, and thus becomes slightly
detuning dependent. After conversion into $\mu$m in object space,
the RMS spot size $\Delta z_{cam}$ of a recorded ion image appears
as a convolution of the thermal spread $\Delta z_{th}$ with the
finite PSF width $\Delta z_{PSF}$. Assuming the PSF to be Gaussian,
we have:

\begin{align}
    \Delta z_{cam}=\sqrt{\Delta z_{PSF}^2+\Delta z_{th}^2}.
    \label{eq:width}
\end{align}

\section{Measurement and results}
\subsection{Time-averaged spatial distribution}

The inset of Fig.~\ref{fig:spreadw} shows an image of a Mg$^+$ ion
trapped with a secular frequency \mbox{$\omega=2\pi\times 15$~kHz}
and cooled close to the Doppler limit of about \mbox{$T\approx1$~mK}
(\mbox{$\overline n\approx 1400$}). From Eq.~\ref{eq:zrms} follows a
RMS spatial spread of \mbox{$\Delta z_{th}\approx 8$~$\mu$m} which
is about  eight times larger than the PSF of our imaging system. In
this case, with Eq.~\ref{eq:width}, the PSF only contributes
$\approx1\%$ to the width $\Delta z_{cam}$ of the image. The laser
detuning is set to \mbox{$\Delta=-2\pi\times
40$~MHz~$\approx-\Gamma$}, and the total laser intensity -- with
both laser intensities equal at the location of the ion -- is
limited to \mbox{$s\lessapprox0.1$} which is monitored by the
observed photon scattering rate. Figure~\ref{fig:spreadw} shows a
histogram of the axial spatial distribution obtained from the image.
No statistically significant deviation from a Gaussian could be
found with a fit, as can be seen from the residuals, which confirms
the stochastic nature of the ion's motion. The RMS spot size of the
ion image of $\Delta z_{cam}=7.7(1)$~$\mu$m
{is indicated by the horizontal bar}, while the vertical bar shows
the resolution of our imaging system.

\subsection{Spatial thermometry measurement}

In order to obtain a precise value of the absolute temperature for
larger oscillation frequencies $\omega$ as well, we need to
determine  $\Delta z_{PSF}$ more accurately. We use the fact that,
according to the laws of Brownian motion, the width $\Delta z_{th}$
of the thermal distribution varies $\propto1/\omega$
(Eq.~\ref{eq:zrms}). From measurements of the spot sizes
\mbox{$\Delta z_{cam}$} for different $\omega$, we obtain
\mbox{$\Delta z_{PSF}$}, \mbox{$\Delta z_{th}$}, and thus the
absolute temperature $T$ of the ion from a fit of Eqs.~\ref{eq:zrms}
and~\ref{eq:width}. This method depends on the shape of
the PSF and the constancy of its width $\Delta z_{PSF}$. The
ion images at high secular frequencies which reflect the PSF do not
show a significant deviation from a Gaussian. To ensure the
constancy of its width, we readjust the imaging system for optimal
resolution at each data point, thus compensating the effects of ion
position changes due to the varying balance between the trap
potentials and the laser force. At the same time we minimize
radial micromotion at each data point to avoid axial heating via
possible coupling to the radial motion.
Figure~\ref{fig:iontemperature}~(a) shows such a measurement for
\mbox{$\Delta=-2\pi\times18.7$~MHz} and $s\lessapprox0.1$ with
$\omega$ between \mbox{$2\pi\times 12$} and $124$~kHz. The insets
show corresponding images and the resulting axial RMS spot sizes
$\Delta z_{cam}$ obtained from Gaussian fits to the image histograms
are plotted versus \mbox{$\omega/2\pi$}. Note that the radial widths only increase slightly by about 15~\% towards
lower axial potentials because radial micromotion compensation
becomes increasingly difficult. The $1/\omega$ behavior of
$\Delta z_{th}$ is confirmed by a fit (Eq.~\ref{eq:width}), which
yields an ion temperature of $T=1.02(3)$~mK and a PSF of $\Delta
z_{PSF}=1.13(3)$~$\mu$m, both with statistical uncertainties.

\subsection{Systematic effects}

The main systematic uncertainty arises from residual axial
micromotion caused by the axial component of the rf fringe fields.
The simultaneous presence of dc and rf fields leads to a parametric
potential $V(t)=\frac{1}{8}m\Omega^2(a-2q\cos(\Omega t))z^2$.
$\Omega$ is the trap frequency. $a\propto eU_{dc}/m\Omega^2$ and
$q\propto eV_{rf}/m\Omega^2$ are trap parameters associated with the
dc and rf voltages $U_{dc}$ and $V_{rf}$, respectively. Without
damping and diffusion, the ion's equation of motion is a Mathieu
equation and for certain parameter sets $(a,q)$ there are stable
trajectories of the ion $z(t)=z_s\cos(\omega
t)(1+\frac{q}{2}\cos{\Omega t})$ with amplitude $z_s$ and
$\omega\approx\frac{1}{2}\Omega(a+q^2/2)^{1/2}$~\cite{micromotion}.
The latter can be interpreted as the secular frequency of the ion in
a time averaged pseudopotential. For vanishing $a$ it reduces to
pure rf confinement $\omega_{rf}=q\Omega/\sqrt{8}$ while pure dc
confinement $\omega_{dc}=\sqrt{a}\Omega/2$ results for $q=0$.
Superimposed on the secular motion, the ion performs micromotion at
the trap frequency $\Omega$ with an amplitude $z_{\mu}=qz_s/2$. For
an ion cooled close to the Doppler limit of 1~mK, $z_{\mu}$ is of
the order of 50~nm and does not influence the apparent spread or
lead to significant deviations from a Gaussian
distribution~\cite{Cirac1994,blatt}.

A stronger effect due to micromotion is expected via the kinetic
energy which is given by~\cite{micromotion}
\begin{equation}
            E_{kin}=\frac{1}{4}m\omega^2z_s^2+\frac{1}{4}m\omega^2_{rf}z_s^2+\frac{1}{2}m\frac{\omega_{rf}^2\omega_{dc}^2}{\omega^2}\Delta
z_0^2.
\label{eq:kinen}
\end{equation}
The first term is due to secular motion, the second reflects
the contribution from the unavoidable {\it ordinary} micromotion
that is associated with the secular motion. It depends on the
secular amplitude $z_s$ and is significant for low secular
frequencies $\omega\approx\omega_{rf}$. The third term is
caused by a mismatch $\Delta z_0$ between the rf and dc potential
minima positions. It results in a displacement $\delta z=\Delta
z_0\omega_{dc}/\omega$ of the ion out of the rf minimum so that it
is exposed to {\it excess} micromotion~\cite{micromotion} that
contributes an energy
$E_{kin}^{excess}=\frac{1}{2}m\omega_{rf}^2\delta z^2$. A cooling
laser introduces both damping and a stochastic force to the equation
of motion.  Blatt et al.~\cite{blatt} have solved the corresponding
Fokker-Planck equation for the first two terms in
eq.~\ref{eq:kinen}, which results in Gaussian spatial and velocity
distributions. It turns out that the presence of a stochastic force
transforms part of the micromotion energy into irregular motion
(temperature) by rf heating which is reflected in a broadening of
the velocity distribution as well as the spatial distribution as
compared to the case of pure secular motion.

From the lowest secular frequency that we have observed when
lowering the dc potential $a$, we estimate an upper limit for the
residual axial rf confinement of $\omega_{rf}=2\pi\times 7$~kHz,
corresponding to $q=8.8\times 10^{-4}$. According to~\cite{blatt}, the
corresponding {\it ordinary} rf heating broadens the spatial
distribution by a factor of 1.15 for the lowest secular frequency of
$\omega=2\pi\times12$~kHz. This broadening decreases strongly
towards higher $\omega\gg\omega_{rf}$. For the estimated upper limit
of $\omega_{rf}=2\pi\times 7$~kHz, our Eq.~\ref{eq:zrms} thus
overestimates the temperature by a factor of 1.3 which results in a
systematic uncertainty of $\lessapprox0.3$~mK in our experiment.

\subsection{Thermometry of a Doppler cooled ion}

In the next step, we employ our thermometry method to investigate
the detuning dependence of Doppler cooling.
Figure~\ref{fig:iontemperature}~(b) shows absolute temperatures
measured for laser detunings $\Delta$ between $-0.2$ and
$-1.0$~$\Gamma$ indicated by squares while the circles are the
Doppler temperatures predicted for pure secular motion by
Eq.~\ref{eq:temperature}. Our measurements follow the detuning
dependence of Eq.~\ref{eq:temperature} qualitatively, but show an
offset of $\approx 0.2$~mK. The higher temperatures are on the order of our conservative estimate for the systematic uncertainty.
However, such a temperature rise can also be caused by {\it excess}
micromotion in a significantly smaller residual axial rf
potential~\cite{micromotion}. While the ordinary micromotion becomes
insignificant for $\omega\gg\omega_{rf}$, excess micromotion due to
a large $\Delta z_0\gg z_s$ leads to a $\omega$-independent
contribution \mbox{$E_{kin}^{excess}\approx
\frac{1}{2}m\omega_{rf}^2\Delta z_0^2$} in Eq.~\ref{eq:kinen} and
may thus increase the ion temperature through rf heating
correspondingly.


\begin{figure}[t]
        \centering
        \includegraphics[width=0.95\columnwidth]{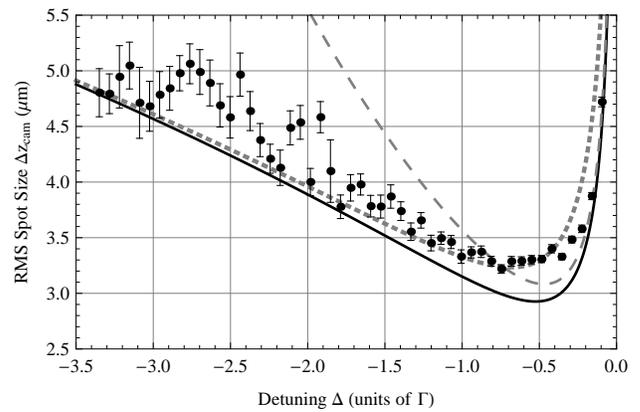}
        \caption{Measured ion image spot size $\Delta z_{cam}$ versus
        laser detuning for $s\lessapprox0.1$ and fixed
        \mbox{$\omega=2\pi\times32$~kHz}. The black solid
        line shows the result expected from Doppler cooling according to
        Eqs.~\ref{eq:temperature}, together with Eq.~\ref{eq:zrms} and
        \ref{eq:width} using $s=0.1$ and $\Delta z_{PSF}=1.5~\mu$m.
        Qualitative agreement with the overall temperature behavior
        of Eq.~\ref{eq:temperature} is found. However quantitatively
        the measured spreads and thus the temperatures are about 10\%
        higher than expected which we attribute to rf heating due to
        excess micromotion, which depends on the photon scattering
        rate and thus decreases towards larger detunings. The gray
        long-dashed line represents Doppler cooling with an additional
        detuning-independent heating rate, the short-dashed line assumes
        a broadening of the line to $1.1~\Gamma$ as caused, e.g., by
        magnetic fields. Both assumptions can not explain the observed
        thermal spread. Larger error bars for larger negative detunings
        are due to lower photon count rate.}
        \label{fig:widthdetuning}
\end{figure}

In order to study the temperature over a wider range of detunings
between 0 and $-3.5~\Gamma$, we measure the image spot sizes $\Delta
z_{cam}$ of the ions for fixed values of \mbox{$\omega=2\pi\times
32$~kHz} and \mbox{$s\lessapprox 0.1$}
(Fig.~\ref{fig:widthdetuning}). At this secular frequency the
ordinary micromotion increases the ion temperature by less than 1\%.
Due to possible variations of the resolution $\Delta z_{PSF}$ during
the laser scan, we do not extract the temperature in this
measurement. Nevertheless, the spot size $\Delta z_{cam}$ can still
serve as a decent measure for the temperature. The solid line in
Fig.~\ref{fig:widthdetuning} represents the behavior according to
Eq.~\ref{eq:temperature} in combination with Eqs.~\ref{eq:zrms}
and~\ref{eq:width}, using $s=0.1$ and $\Delta z_{PSF}=1.5$~$\mu$m.
The latter is estimated to be an upper limit. Quantitatively, the
measured values are again up to 10\% higher. The data supports our
hypotheses that rf heating due to excess micromotion is the main
cause of the higher observed temperatures. Since rf heating is
related to the photon scattering of the cooling laser~\cite{blatt},
its rate is expected to be detuning-dependent similar to the cooling
rate. The wide tuning range in this measurement allows to exclude
two other possible effects as causes for the excess temperature. As
a comparison, the long-dashed line shows the expected broadening
taking an additional, detuning-independent heating rate into
account. The short-dashed line assumes that the line width is
homogeneously broadened to $1.1~\Gamma$ caused, e. g., by the Zeeman
splitting or micromotion-induced line broadening~\cite{micromotion}.
From earlier spectroscopy experiments~\cite{maxmg}, we know that for
our experimental parameters the homogeneous broadening is in fact
lower. In conclusion, both effects can be excluded as causes for the
observed higher thermal spreads (temperatures).\\

\section{Summary}

To summarize, we have demonstrated that a single, Doppler cooled ion
weakly bound in an ion trap, shows a Gaussian thermal spatial
distribution as expected from Brownian motion. We show that this
time-averaged spatial distribution can be used for sub-milliKelvin
thermometry with an accuracy of $<0.3$~mK unchallenged by similar
methods in the unresolved sideband regime. Further, we have
employed this thermometry to investigate the detuning dependence of
the temperature in Doppler cooling. Note that this method which was
demonstrated for the axial trapping direction here, can in principle
be applied to any projection of an ion on the imaging plane. The use
of high NA objectives with resolutions close to the diffraction
limit~\cite{kielpinskiimaging,kielpinski} would allow to further
improve on accuracy and to perform thermometry in steeper trapping
potentials.

\end{document}